\documentclass[12pt,preprint]{aastex}
\usepackage{lscape}

\shorttitle{Seismic Test of Solar Models}
\shortauthors{Winnick et al.}

\begin{document}

\title{Seismic Test of Solar Models, Solar Neutrinos and Implications
for Metal-Rich Accretion}

\author{R. A. Winnick, Pierre Demarque and Sarbani Basu}
\affil{Department of Astronomy, Yale University, Box 208101, 
New Haven, CT 06520-8101}
\and
 \author{D. B. Guenther}
\affil{Department of Astronomy and Physics, Saint Mary's University, 
Halifax, NS, Canada, B3H 3C3}

\begin{abstract}

The Sun is believed to have been the recipient of a substantial amount
of metal-rich material over the course of its evolution, particularly in
the early stages of the Solar System.  With a long diffusion timescale, 
the majority of this accreted matter should still exist in the solar
convection zone, enhancing its observed surface abundance, and 
implying a lower-abundance core.  While helioseismology rules out 
solar models with near-zero metallicity cores, some solar models with 
enhanced metallicity in the convection zone might be viable, as small 
perturbations to the standard model.  Because of the reduced interior 
opacity and core temperature, the neutrino flux predicted for such 
models is lower than that predicted by the standard solar model.  This
paper examines how compatible inhomogeneous solar models of this kind 
are with the observed low and intermediate degree {\it p}-mode 
oscillation data, and with the solar neutrino data from the SNO 
Collaboration.  We set an upper limit on how much metal-rich accretion
took place during the early evolution of the Sun at $\sim$ 2M$_\oplus$
of iron (or $\sim$ 40M$_\oplus$ of meteoric material). 
 
\end{abstract}


\section{Introduction}

Early on in the evolution of the Sun, the interstellar medium and young 
Solar System provided an environment capable of frequent accretion events.  
Material in the form of ISM, proto-planets, planetesimals, comets, and
asteroids frequently bombarded the Solar surface and were accreted.  This
accreted matter is deficient in hydrogen and helium, introducing metal-rich
material into the upper layers of the Sun.  Exactly how much metal-rich
matter was accreted in these early stages and for how long, has been 
speculated by several authors (Christensen-Dalsgaard, Gough, \& Morgan
1979, Jeffery et al. 1997), but unfortunately is not well known. 

More recently, Murray et al. (2001) have searched a sample of 640
solar-type stars for the signature of iron enhancement in their spectra 
and concluded that, on average, these stars appear to have accreted 
about 0.5M$_\oplus$ of iron while on the main sequence.  They raise the 
possibility that the Sun may have accreted a similar amount of iron 
during its evolution.  Whether helioseismology can detect this amount of 
accretion is uncertain.  So far only one study has been published on 
the effect of accretion on the {\it p}-mode frequency (Henney \& Ulrich 1998).
These authors concluded that the accretion of 8M$_\oplus$ 
of meteoric material (approximately 0.4M$_\oplus$ of iron) on the
Sun could not be detected by seismology due to other uncertainties in
the models.

The main consequence of such accretion is the metal enrichment of the 
Sun's surface and convection zone, while maintaining a lower-abundance
core.  In such a scenario, the expected neutrino flux is reduced due
to the lower opacity and temperature in the central region of the Sun.  
It has long been known that solar models with near-zero metallicity 
cores can lower the predicted neutrino flux, and this explanation was 
discussed early-on as a possible solution of the classical neutrino 
problem (see e.g. the review by Rood 1978).  More recently, Guenther 
\& Demarque (1997, hereafter GD97) have constructed solar models with 
low-Z cores using present physics and found their {\it p}-mode frequencies
to be incompatible with solar {\it p}-mode observations.

This paper describes work which can be viewed as an extension of GD97;
but this time we examine the limits set by helioseismology on the 
non-standard assumption that the solar convection zone of the Sun has 
been enriched in heavy elements by accretion during its early
evolution.  In addition to comparing the {\it p}-mode frequencies to
observation in the frequency difference diagram, as done by GD97, we 
also compare the calculated sound-speed and density in our models to
the same quantities derived from observation by inversion (Basu \&
Antia 1997, hereafter BA97; Basu, Pinsonneault \& Bahcall 2000, 
hereafter BPB00).

Since this research was begun, the first results from 
the Sudbury Neutrino Observatory (SNO) experiment have been released 
(Ahmad et al. 2001).  Ahmad et al. (2001) conclude that 
the $^{8}$B neutrino flux derived from their observations, i.e.
$\Phi_{^{8}B}$ = 5.44 $\pm$ 0.99 $\times$ $10^{6}$ cm$^{-2}$s$^{-1}$, 
agrees with the predictions of the best standard solar models.  Any 
accretion model must then also be constrained by the SNO observations,
in addition to the constraints of helioseismology.

\section{Assumptions and Method}

The main assumption in this paper is that the accretion of heavy 
elements onto the Sun took place during the early evolution of the
Sun, at a time when the Sun was near the main sequence,
i.e.  when the convection zone had reached its main sequence value.
The convection is shallow at this time, and only the outer  
2\% of the Sun's mass in the convection zone get enriched.  

As in GD97, the solar models were 
constructed using the Yale stellar evolution code (YREC for Yale Rotating 
Evolution Code) in its non-rotating configuration (Guenther et al. 1992).  
The energy generation routines used in YREC are those of Bahcall \& 
Pinsonneault (1992).  The neutrino cross sections used are the same as
in Bahcall, Pinsonneault \& Basu (2001).  The equation of state tables
prepared by the OPAL researchers (Rogers 1986, Rogers, Swenson, \& 
Iglesias 1996) were used for the models.  The interior opacities were
derived from the OPAL opacity tables (Iglesias \& Rogers 1996), while
the surface and atmospheric opacities were taken from the Alexander 
\& Ferguson (1994) tables.

All models were evolved from a zero-age main sequence (ZAMS) model to
near the age of the Sun in 50 equally spaced time steps.  The age of 
the Sun was taken to be 4.5 Gyr, close to the meteoric solar age 
(Guenther 1989).  As was noted by GD97, standard solar models of this
age are in best agreement with the calculated oscillation spectra.  
The mixing length/pressure scale height ratio, $\alpha$, and the 
helium content by mass, Y, were adjusted automatically by the code to
produce a precisely calibrated standard solar model (Guenther et al. 
1992).  Most models were tuned to match the 
observed solar radius to one part in $10^7$ (R$_\odot$=6.958 x $10^{10}$cm) 
and the solar luminosity to one part in $10^6$ (L$_\odot$=3.8515 x 
$10^{33}$ergs s$^{-1}$).  Models \#15, 16, and 19 (see Table 1) did not 
converge at the aforementioned tolerance for radius, but converged to 
the solar radius to one part in $10^6$.
The number of shells in standard solar models \#17 and 18 was 1892, and 
1893 in models \#19 and 20.  There were more than 2300 shells in all 
non-standard solar models.  Tests have shown that at least that many
shells are needed to obtain the required precision in the calculation of the 
{\it p}-mode frequencies (Guenther et al. 1992; Guenther 1994).   

Helium and heavy element diffusion were included in all of the solar 
models due to their necessary role in finding simultaneous agreement with 
both the observed value of (Z/X)$_\odot$ and the {\it p}-mode spectrum 
(for further details see GD97).  The assumption made regarding the 
nonstandard models is that due to the accretion of metal-rich matter, 
the Sun's interior, (defined here as R $<$ R$_{\mathrm {env}}$), is 
composed of material at a lower heavy-element abundance than the Sun's
surface.  In order to produce such nonstandard solar models with low-Z
interiors, the run of Z in the initial ZAMS model was modified.  The 
interior metal abundance was initially set to the homogeneous value of
Z$_{\mathrm {int}}$ out to (M/M$_\odot$)=0.9, indicating a metal-poor 
interior.  The more metal-rich exterior of the Sun, 
(M/M$_\odot$)$\ge$0.975, was setup with a metal abundance of 
Z$_{\mathrm {init}}$.  Z$_{\mathrm {int}}$ and Z$_{\mathrm {init}}$ 
represent ZAMS mass fractions of all heavy elements for the interior
and exterior, respectively.  In the intermediate region, 
0.9$\le$(M/M$_\odot$)${\le}$0.975, Z linearly increases from 
Z$_{\mathrm {int}}$ to Z$_{\mathrm {init}}$.

In deciding what value Z$_{\mathrm {int}}$ might be for the nonstandard 
solar models, it is relevant to note that the Sun is observed to be
more metal rich than the surrounding ISM, with Z$_{\mathrm {ISM}}$ 
possibly as low as 65\% of Z$_\odot$ (Mathis 1996).  Believing the 
Sun formed from material typical of the ISM, the metal-enhanced
exterior could have resulted from the bombardment of metal-rich 
material in the form of comets, asteroids, planetesimals and 
proto-planets.  With a long diffusion timescale, most of this material
should still exist in the upper layers of the Sun, leaving the
interior metal-poor, much like the surrounding ISM.  With this in
mind, we chose a value of Z$_{\mathrm {int}}$=0.65Z$_\odot$.  Other 
values of 0.30Z$_\odot$, 0.50Z$_\odot$, and 0.80Z$_\odot$ were also 
examined. 

A choice of Z$_{\mathrm {int}}$ = 0.80Z$_{\mathrm {init}}$ corresponds to 
an accretion enhancement of about 2M$_\oplus$ in iron, or about 
40M$_\oplus$ in meteoric material.  Similarly, 
Z$_{\mathrm {int}}$=0.65Z$_{\mathrm {init}}$, 
Z$_{\mathrm {int}}$=0.50Z$_{\mathrm {init}}$, and 
Z$_{\mathrm {int}}$=0.30Z$_{\mathrm {init}}$ corresponds to about 
2.9M$_\oplus$, 4.1M$_\oplus$ and 5.7M$_\oplus$, respectively, in iron 
accretion, or about 60M$_\oplus$, 80M$_\oplus$ and 120M$_\oplus$,
respectively, in meteoric material accretion.

Four standard solar models were created, differing only by the value of
Z$_{\mathrm {init}}$, the initial or ZAMS mass fraction value of all
heavy elements in the solar exterior.  Without the assumption of a
more metal-poor interior, Z$_{\mathrm {int}}$ is assumed equivalent to 
Z$_{\mathrm {init}}$.  Values of Z$_{\mathrm {init}}$ were taken as
0.0170, 0.0188, 0.0200, and 0.0220.  These standard solar models 
comprise models \#17-20 in Tables 1-3.  Z$_{\mathrm {init}}$ 
was also varied in the nonstandard solar models, thus along with the 
varying Z$_{\mathrm {int}}$, a grid was created, comprising models 
\#1-16 in Tables 1-3.

Physical characteristics of both the standard (\#17-20) and 
nonstandard (\#1-16) solar models are listed in Table 1.  
Table 1 includes, from left to right:  Model, the model number;
Type, the type of model, where NSSM stands for a nonstandard solar 
model and SSM stands for a standard solar model; X$_{\mathrm {init}}$,
the initial or ZAMS mass fraction value of hydrogen; 
Z$_{\mathrm {init}}$, the initial or ZAMS mass fraction value of all 
heavy elements in the solar exterior; 
Z$_{\mathrm {int}}$/Z$_{\mathrm {init}}$, the initial or ZAMS mass
fraction value of all heavy elements in the solar interior relative
to Z$_{\mathrm {init}}$; X$_{\mathrm {surf}}$, the surface mass 
fraction value of hydrogen at the evolved age; Z$_{\mathrm {surf}}$, 
the surface mass fraction of all heavy elements at the evolved age; 
M$_{\mathrm {env}}$, the fraction of the total mass contained in the 
outer convective envelope; R$_{\mathrm {env}}$, the radius fraction of
the base of the convective envelope; $\log{P_c}$, the base ten
logarithm of the central pressure; $\log{T_c}$, the base ten logarithm
of the central temperature; $\log{\rho_c}$, the base ten logarithm of 
the central density; X$_c$, the central mass fraction of heavy
elements; Z$_c$, the central mass fraction of heavy elements.

The nuclear energy generation properties of both the standard
(\#17-20) and nonstandard (\#1-16) solar models are listed in Table 2.
Table 2 lists the fraction of total photon luminosity coming
from the PP I, PP II, and PP III branches of the PP network and from 
the CNO cycle.  Also listed are the individual neutrino fluxes from 
the neutrino producing reactions that occur in the sun (see Figure 1 of
GD97).  Note that in GD97 the neutrino fluxes are listed in units of 
10$^{10}$cm$^{-2}$s$^{-1}$.  Finally, Table 2 lists 
${\Phi}$($^{37}$Cl), the total neutrino flux, in SNU, for the 
$^{37}$Cl detector; and ${\Phi}$($^{71}$Ga), the total neutrino flux, 
in SNU, for the $^{71}$Ga detector.

\section{{\it p}-mode Frequencies}

As a test on the validity of a model, the expected oscillation 
frequencies can be compared with observations from the Sun.
Guenther's non-radial, non-adiabatic pulsation program (Guenther 1994) 
was used to calculate the oscillation frequencies of the models 
produced with YREC.  The model output was then compared with data 
obtained by the Michelson Doppler Imager (MDI) instrument on board
the {\it Solar and Heliospheric Observatory (SOHO)} during the first
year of its operation (Rhodes et al. 1997).  This data set was chosen
as it comprises one of the longest time-series of data, 360 days.  
More recent MDI-SOHO data include only 144-day or 72-day data sets, and
therefore restrict the number of data points, with most sets having
very few {\it l}=0,1,2 modes.  Data from the GOLF experiment (see 
Bertello et al. 2000 and Garc\'\i a et al. 2001 for the latest
results) were not considered here despite the low-degree low-frequency
modes.  While these modes can tell us a lot about the structure of 
the core (see Turck-Chi\`eze et al. 2001), for our purposes here, it 
requires mixing one set of low degree modes with intermediate and 
high degree modes from another instrument, which can give rise to 
artifacts in the solar core unless the data are contemporaneous (see 
Basu et al. 1996, 1997).  Hence, in an attempt to avoid such
artifacts, we have opted for a homogeneous set of data.
For comparison, BiSON+LOWL (Basu et al. 1997) data were also used.
This data set had been specially constructed by obtaining frequencies 
from contemporaneous BiSON and LOWL time series. This set gave results
very similar to those obtained by the MDI-SOHO data set used here.

From these data, {\it p}-mode frequency differences:  
$\nu_{\mathrm {model}}$-$\nu_\odot$, were computed for {\it l}=0-4, 10, 
20, 30, 40, 50, 60, 70, 80, 90, and 100.  Frequency difference plots 
(model frequency minus observed frequency versus observed frequency)
were then constructed for each model and can be seen in Figure 1.
Each line joins together data of a common {\it l}-value, hence joining
together {\it p}-modes with approximately similar inner turning
points.  In the event of perfect seismic agreement with the Sun, a 
frequency difference plot would show a straight horizontal line at 0 
$\mu$Hz, indicating that the {\it p}-mode frequency differences were 
zero.  Instead, Figure 1 shows a more complex bundle of lines, 
indicating discrepancies of our models from the Sun.  As in GD97, the 
quality of a models' agreement with the Sun is judged on the tightness
of the bundle of lines.  The underlying slope error present is thought
to be due to modeling errors in the very outermost layers of the Sun, 
a region of known uncertainties.  The errors in the interior of the 
model, primarily near the base of the convection zone, are directly 
correlated to the bundle thickness in the sense that the tighter the 
bundle, the better the model is at fitting the region near and above 
the convection zone.  (See GD97 for further details.)

Figure 1 is a 4$\times$5 grid of plots showing the frequency 
differences for each of the 20 models computed.  Each plot is
annotated by its model number (see Tables 1-3), the surface 
Z/X ratio, and $^{8}$B neutrino flux, in cm$^{-2}$s$^{-1}$.  Lines 
connect common {\it l}-values, with {\it l}=0-4, 10, 20, 30, 40, 50, 
60, 70, 80, 90, and 100.  As indicated by the model numbers, the four 
standard solar models computed in this work comprise the last row in 
Figure 1.  Figure 2 shows a larger version of the frequency-difference
plot for model \#20, including error bars from the MDI-SOHO data.  The
error bars indicate 10$\sigma$ errors in the data averaged over 
250$\mu$Hz frequency bins, for all {\it l}-values included in this 
work.  As can be seen in Figure 2, the errors increase with increasing
frequency.  Despite the relatively good agreement of our model with
the Sun, the accuracy of the observations is such that our best model 
is discrepant at roughly the 30$\sigma$ level at low frequencies, and 
is worse at higher frequencies.  This large discrepancy seen in the 
calculated frequencies is not found in the sound speed comparison 
(see Figure 6).  It is primarily due to modeling errors in the surface
layers of the Sun, which could be reduced by further analyses
including magnetic fields, turbulence, and a better understanding of 
the convection zone.

Special consideration was given to low {\it l p}-modes due to their 
ability to probe the deep interior of the Sun.  This is due to the 
fact that the inner turning points of these modes are located closest 
to the core.  Low {\it l p}-modes are still sensitive to the outer 
layers, but this effect
can be canceled out by subtracting from a given {\it p}-mode frequency
the frequency of a {\it p}-mode with a similar eigenfunction shape in 
the outer layers and distinct eigenfunction shape in the deeper
layers.  The small spacing difference, defined as 
$\delta\nu({\it n,l})=\nu({\it n,l})-\nu({\it n}-1,{\it l}$+2), 
thus provides a further diagnostic of the deep interior of the Sun.  
(See GD97 for further details.)  To compare our models with 
observations from the Sun, small spacing differences plots were 
produced [i.e., 
$\delta\nu_{\mathrm {model}}({\it n,l})-\delta\nu_\odot({\it n,l})$ 
{\it vs.} $\nu_\odot({\it n,l})$].  As in the case of the frequency 
difference plots, the closer 
$\delta\nu_{\mathrm {model}}({\it n,l})-\delta\nu_\odot({\it n,l})$ 
is to 0 $\mu$Hz, the better the agreement between the observations and
a model.

Since the sensitivity of the small spacings to the deep interior of 
the Sun diminishes with increasing {\it l}, only those {\it p}-modes 
with {\it l}=0, 1, and 2 were used to contrast the models with
observed values from the Sun.  The small spacing difference plots can
be seen in Figure 3.  Figure 3 is a 3$\times$5 grid in which each row 
contains models of the same Z$_{\mathrm {int}}$, where 
Z$_{\mathrm {int}}$=0.30Z$_{\mathrm {init}}$ includes models \#1-4, 
Z$_{\mathrm {int}}$=0.50Z$_{\mathrm {init}}$ includes models \#5-8, 
Z$_{\mathrm {int}}$=0.65Z$_{\mathrm {init}}$ includes models \#9-12, 
Z$_{\mathrm {int}}$=0.80Z$_{\mathrm {init}}$ includes models \#13-16, 
and Z$_{\mathrm {int}}$=Z$_{\mathrm {init}}$ includes the standard solar 
models \#17-20.  Each column in Figure 3 contains results for a common
{\it l}-value, with {\it l}=0 in column 1, {\it l}=1 in column 2, and 
{\it l}=2 in column 3.  Lines in each panel represent models with the 
same Z$_{\mathrm {int}}$ with differing Z$_{\mathrm {init}}$ values
ranging from 0.0170 to 0.0220, as indicated.

\section{Sound-speed and Density}
 
Figure 4 shows the relative sound-speed differences for the twenty 
models presented in this paper compared to the observed solar 
sound-speed, derived from an inversion of the solar {\it p}-mode 
frequencies (BPB00).  We note the superior agreement of the standard 
solar models with observation, especially model \#20, as compared to 
the non-standard models.  Similarly, Figure 5 shows the same plots for
density.  The less precise agreement with densities than with 
sound-speed is in part a reflection of the greater uncertainty in 
density inversions.

\section{Solar Models}
\subsection{Choice of a Benchmark Standard Solar Model}

Figure 1 shows the characteristics of the standard solar models 
(\#17-20) in the frequency difference diagram. Best agreement with 
observation, as measured by the thickness of the line bundles 
corresponding to different $l$-values, favors \#20 and 19.  Note 
that our model \#19 is nearly identical to model \#20 of GD97, that 
was judged by GD97 as the ``best'' standard solar model in their study. 
But our model \#20, which is intermediate in Z$_{\mathrm {init}}$
between GD97's models \#20 and 21, is a better model still.  The 
bundle of $l$-values in model \#20 is thinner than in model \#19, a 
fact which is also reflected in the best agreement with observation in
the sound-speed plot shown in Figure 4.

In selecting a ``best'' standard solar model, one should also take into 
account two additional constraints.  The first constraint, to which we
will assign the most weight, is the solar convection zone depth, 
R$_{\mathrm {env}}$, which has been derived by sound-speed inversion 
(Christensen-Dalsgaard et al. 1991; BA97).  The second additional 
constraint is Y$_{\mathrm {surf}}$ for the Sun.  This quantity has also 
been estimated by inversion of the {\it p}-mode data (Basu 1998).  

For model \#20, we have R$_{\mathrm {env}}$ = 0.7128, to be compared to 
R$_{\mathrm {env}}$ = 0.713 $\pm$ 0.001 from inversion (BA97). This 
agreement is confirmed by the sound-speed plot for model \#20 (see 
Figures 4 and 7), which shows close agreement with the inversion data.
At the same time, model \#20 yields Y$_{\mathrm {surf}}$ = 0.2519, to 
be compared to the helium mass fraction 0.248 $\pm$ 0.003 derived by 
Basu (1998), a satisfactory agreement in view of the uncertainties in 
diffusion coefficients and efficiency.  In addition, the $^{8}$B 
neutrino flux of 5.63 cm$^{-2}$s$^{-1}$ for model \#20 is in excellent
agreement with 
$\Phi_{^{8}B}$ = 5.44 $\pm$ 0.99 $\times$ $10^{6}$ cm$^{-2}$s$^{-1}$ 
found by SNO (Ahmad et al. 2001).  On the other hand, the surface 
metallicity (Z/X)$_{\mathrm {surf}}$ is a little too large.  One can 
see that a model interpolated between models \#20 and 19 would satisfy
all constraints within the errors.  

As it is, model \#20 is an excellent standard solar model, as 
illustrated in Figures 4-8. Comparing Figures 2 and 3 in 
BPB00 with our Figures 7 and 8, we see that model \#20 compares 
favorably with model BP2000.

The small spacing plots (Figure 3) shows that all our standard models,
models \#17-20, agree equally well with observation.  Within the range
of chemical composition we have considered, small spacings are 
insensitive to the choice of initial heavy element content.  Figure 6
shows the zoomed-in small spacing plot for $l$=0 for standard solar 
model \#20.  Error bars calculated from the MDI-SOHO data set used 
are also plotted.  The observational uncertainties are approximately 
$\pm$0.07 $\mu$Hz for the small spacing determination.  From this
figure, the numerical uncertainties in the solar model appear to be 
approximately $\pm$0.05 to 0.10 $\mu$Hz for the small spacing 
calculation.  The larger uncertainties of $\pm$1 $\mu$Hz for the mode 
frequencies themselves cancel out somewhat when calculating the 
small spacings.

It is difficult to evaluate the uncertainty in the observed solar 
heavy element content (Z/X)$_{\mathrm {surf}}$.  GD97 used 
(Z/X)$_{\mathrm {surf}}$ = 0.0244 $\pm$ 0.001, from Grevesse, Noels 
\& Sauval (1996).  More recent lower estimates suggest that large 
systematic errors may still exist in the measurements.  Grevesse \& 
Sauval (1998) find (Z/X)$_{\mathrm {surf}}$ = 0.0230, quoting a 10\% 
error estimate.  Asplund (2000), using a stellar model atmosphere 
constructed with a more realistic treatment of convection, derives 
(Z/X)$_{\mathrm {surf}}$ = 0.0226.  Unfortunately, theoretical estimates
for both  Y$_{\mathrm {surf}}$ and  (Z/X)$_{\mathrm {surf}}$ are also 
difficult to make, due to the uncertainties in calculating the 
efficiency of diffusion just below the convection zone in the stellar 
interior (Chaboyer et al. 1995). Helioseismic sound-speed inversion 
shows that the tachocline structure is one of the least well-understood
features of standard solar models (BA97). 

\subsection{Non-Standard Solar Models}

At first glance, the frequency difference diagram in Figure 1 
does not reveal much difference in bundle thickness between the 
non-standard models \#7, 8, 11, and 15 and the standard model 
\#19.  The $l$-value bundle for model \#12 seems even slightly
thinner.  But models \#7, 8, 11 and 12 are all readily ruled 
out by the $^{8}$B neutrino constraint.

Non-standard model \#16 is similar to standard model \#20 in having
the thinnest $l$-value bundles and a marginal value for 
(Z/X)$_{\mathrm {surf}}$.  But it fails the $^{8}$B neutrino 
test by a small amount.  Non-standard models \#10 and 15, like 
standard model \#20, pass the convection zone depth test, but of 
these, only \#15 passes the (Z/X)$_{\mathrm {surf}}$ test.  However,
model \#15 fails the $^{8}$B neutrino test.

Turning to the small spacing diagram (Figure 3), we see that the
non-standard models agree as well with observation as the standard 
models only for $l$ = 0.  But for $l$ = 1 and $l$ = 2, there is an 
increasing discrepancy with decreasing Z$_{\mathrm {int}}$.  The 
maximum discrepancy is only 2 $\mu$Hz, however, which may still be 
within the uncertainties.  As in the case of the standard models, 
heavy element content plays little role.

As a further comparison of the models calculated here, the 
root-mean-square of the sound-speed difference, rms($\delta$c/c), 
was computed for each model.  Seen in Table 3, the results indicate 
that those models with Z$_{\mathrm {init}}$=0.0220 fair well.  Model 
\#20 certainly outperforms the others, followed by standard solar 
model \#19 and non-standard solar model \#16.  These results further 
strengthen the argument that model \#20 is in best agreement.
Table 3 summarizes the values of the constraints considered here
for each solar model computed in this work.  Listed are the
Model, R$_{\mathrm {env}}$, Y$_{\mathrm {surf}}$, $^{8}$B, 
(Z/X)$_{\mathrm {surf}}$ and rms($\delta$c/c).

We conclude that of all the non-standard solar models listed in 
Tables 1 and 2, only models \#15 and 16 are marginally acceptable.  
We can use these two models to set an upper limit to the amount of 
heavy element accretion during the early main sequence evolution of 
the Sun.  

\section{Discussion}

There has been interest during the past few years in the effects of
accretion on the Sun during its evolution.  With this application in 
mind, the main purpose of this paper was to study the properties of a 
number of non-standard solar models in which the convection zone is 
richer in heavy elements than the interior, and to explore the 
constraints of helioseismology on such models.

A set of standard solar models were first constructed using the same 
physics and stellar evolution code as GD97, to serve as benchmarks
for our non-standard models. Although our purpose was not to achieve 
the best possible fit to observation, we find that our standard solar 
model \#20 satisfies the most stringent constraints from 
helioseismology at least as well as the best published standard solar 
model.  It is just outside the 1$\sigma$ limit for the 
(Z/X)$_{\mathrm {surf}}$ constraint, but agrees very well with the 
SNO Collaboration $^{8}$B neutrino flux.  It is apparent that a
standard solar model in full agreement with all up-to-date
observational data would be achieved with the same physics as in GD97,
for a Z$_{\mathrm {init}}$ intermediate between model \#19 and model 
\#20, i.e. for 0.020$<$Z$_{\mathrm {init}}<$0.022.  It would also be 
interesting to probe intermediate values of Z$_{\mathrm {int}}$, with 
0.80$<$Z$_{\mathrm {int}}<$1.00.

For this work, we have only considered Z accretion from cometary 
material after the pre-MS phase, but what about the possibility of 
helium accretion as well, and increased accretion during an earlier, 
pre-MS era?  If most of the accretion takes place in the pre-MS phase 
of solar evolution, we must consider two phases.  In the early phase, 
the proto-Sun is fully or nearly fully convective (the Hayashi phase).  
Since the accreted material is mixed efficiently in the convective 
region, composition gradients in the deep interior would not be expected.  
In the later phase of pre-MS evolution (the Henyey phase), the 
radiative core gets progressively larger until the main sequence is 
reached.  This phase of evolution, which takes place on a thermal 
timescale, is relatively short compared to diffusion timescales, and 
could leave behind a composition gradient in the interior, below the 
convection zone.  Since we do not understand the region
just below the present convection zone well, we cannot rule out the 
possibility of some composition in-homogeneity in the outer radiative 
envelope due to late pre-MS accretion.  However, two factors argue 
against the existence of significant composition gradients in the
outer envelope: (1) the presence of shear-induced turbulence in the 
tachocline region, and (2) the close agreement of the observed 
sound-speed and the SSM sound-speed in the radiative region below the 
tachocline layer. 

None of our non-standard solar models agree as well with observation
as our best standard models.  But two of the least extreme non-standard 
solar models, models \#15 and 16, come close to satisfying all of the 
observational constraints. For this reason, they provide a realistic 
upper limit of $\sim$2M$_\oplus$ of iron (or $\sim$40M$_\oplus$
of meteoric material) to the accretion experienced by the Sun during
its early main sequence phase.  This conclusion is compatible with 
Henney \& Ulrich's (1998) earlier null result that the accretion of 
8M$_\oplus$ of meteoric material on the Sun could not at this 
point be detected by seismology because of the uncertainties. 
Similarly, a solar enrichment of 0.5M$_\oplus$ in iron, as suggested
recently by Murray et al. (2001), is beyond detectability by seismic 
means at the present time.  

\acknowledgements
We are indebted to Prof. J. N. Bahcall for his advice and help with 
the neutrino cross sections.  This work was supported in parts by NASA
grant NAG5-8406 (PD), a grant from NSERC Canada (DBG), and NASA grant 
NAG5-10912 (SB).  This work utilizes data obtained by the Solar 
Oscillations Investigation / Michelson Doppler Imager on the Solar
and Heliospheric Observatory (SOHO). SOHO is a project of
international cooperation between ESA and NASA.  MDI is 
supported by NASA grants NAG5-8878 and NAG5-10483 to Stanford 
University.

\clearpage

\begin{landscape}
\begin{deluxetable}{cccccccccccccc}
\tabletypesize{\scriptsize}
\tablecaption{Solar Model Characteristics}
\tablewidth{0pt}
\tablehead{
\colhead{Model} & 
\colhead{Type} & 
\colhead{X$_{\mathrm {init}}$} & 
\colhead{Z$_{\mathrm {init}}$} & 
\colhead{Z$_{\mathrm {int}}$/Z$_{\mathrm {init}}$} & 
\colhead{X$_{\mathrm {surf}}$} & 
\colhead{Z$_{\mathrm {surf}}$} & 
\colhead{M$_{\mathrm {env}}$}  & 
\colhead{R$_{\mathrm {env}}$} & 
\colhead{$\log{P_c}$}  & 
\colhead{$\log{T_c}$} & 
\colhead{$\log{\rho_c}$} & 
\colhead{X$_c$} & 
\colhead{Z$_c$}
}
\startdata
1  &NSSM &0.8145 &0.0170 &0.30 &0.8359 &0.0153 &0.02010 &0.7139 &17.354 &7.160 &2.148 &0.468 &0.0053 \\
2  &NSSM &0.8066 &0.0188 &0.30 &0.8282 &0.0170 &0.02201 &0.7071 &17.356 &7.162 &2.151 &0.460 &0.0059 \\
3  &NSSM &0.8013 &0.0200 &0.30 &0.8231 &0.0181 &0.02332 &0.7026 &17.357 &7.163 &2.153 &0.456 &0.0063 \\
4  &NSSM &0.7930 &0.0220 &0.30 &0.8152 &0.0199 &0.02474 &0.6982 &17.358 &7.165 &2.155 &0.448 &0.0069 \\
5  &NSSM &0.7813 &0.0170 &0.50 &0.8063 &0.0153 &0.02063 &0.7164 &17.361 &7.170 &2.160 &0.431 &0.0089 \\
6  &NSSM &0.7702 &0.0188 &0.50 &0.7955 &0.0169 &0.02262 &0.7102 &17.364 &7.173 &2.165 &0.420 &0.0099 \\
7  &NSSM &0.7629 &0.0200 &0.50 &0.7884 &0.0180 &0.02408 &0.7057 &17.366 &7.175 &2.168 &0.413 &0.0105 \\
8  &NSSM &0.7546 &0.0220 &0.50 &0.7805 &0.0198 &0.02531 &0.7020 &17.367 &7.178 &2.170 &0.405 &0.0115 \\
9  &NSSM &0.7609 &0.0170 &0.65 &0.7880 &0.0153 &0.02081 &0.7185 &17.366 &7.177 &2.168 &0.408 &0.0116 \\
10 &NSSM &0.7517 &0.0188 &0.65 &0.7791 &0.0169 &0.02251 &0.7130 &17.367 &7.180 &2.171 &0.399 &0.0128 \\
11 &NSSM &0.7454 &0.0200 &0.65 &0.7730 &0.0180 &0.02365 &0.7094 &17.368 &7.181 &2.173 &0.393 &0.0137 \\
12 &NSSM &0.7355 &0.0220 &0.65 &0.7636 &0.0198 &0.02488 &0.7061 &17.370 &7.184 &2.176 &0.384 &0.0150 \\
13 &NSSM &0.7461 &0.0170 &0.80 &0.7751 &0.0153 &0.02058 &0.7211 &17.368 &7.182 &2.172 &0.392 &0.0143 \\
14 &NSSM &0.7357 &0.0188 &0.80 &0.7649 &0.0169 &0.02227 &0.7157 &17.370 &7.185 &2.176 &0.382 &0.0158 \\
15 &NSSM &0.7284 &0.0200 &0.80 &0.7578 &0.0180 &0.02339 &0.7124 &17.371 &7.187 &2.178 &0.374 &0.0168 \\
16 &NSSM &0.7170 &0.0220 &0.80 &0.7470 &0.0198 &0.02474 &0.7088 &17.373 &7.191 &2.182 &0.363 &0.0185 \\
17 &SSM  &0.7265 &0.0170 &1.00 &0.7577 &0.0152 &0.02041 &0.7244 &17.369 &7.190 &2.178 &0.364 &0.0179 \\
18 &SSM  &0.7140 &0.0188 &1.00 &0.7456 &0.0169 &0.02202 &0.7199 &17.371 &7.194 &2.182 &0.349 &0.0198 \\
19 &SSM  &0.7057 &0.0200 &1.00 &0.7376 &0.0180 &0.02325 &0.7163 &17.372 &7.197 &2.185 &0.339 &0.0210 \\
20 &SSM  &0.6960 &0.0220 &1.00 &0.7283 &0.0198 &0.02449 &0.7128 &17.373 &7.201 &2.188 &0.327 &0.0232 \\
\enddata
\end{deluxetable}
\end{landscape}

\begin{landscape}
\begin{deluxetable}{cccccccccccccccc}
\tabletypesize{\scriptsize}
\tablecolumns{16}
\tablecaption{Solar Model Nuclear Data\tablenotemark{a}}
\tablewidth{0pt}
\tablehead{
\colhead{}  & \multicolumn{4}{c}{Fraction of Total Luminosity} 
& \colhead{} & \multicolumn{10}{c}{Solar Neutrino Flux\tablenotemark{b}} \\
\cline{2-5} \cline{7-16} \\
\colhead{Model\#}  & \colhead{PPI}       & \colhead{PPII} & 
\colhead{PPIII}    & \colhead{CNO}       & \colhead{} &
\colhead{PP} & 
\colhead{PeP}      & \colhead{HeP}       & \colhead{$^{7}$Be} & 
\colhead{$^{8}$B}  & \colhead{$^{13}$N}  & \colhead{$^{15}$O} & 
\colhead{$^{17}$F} & \colhead{$\Phi$($^{37}$Cl)} & 
\colhead{$\Phi$($^{71}$Ga)} \\
\colhead{}         & \colhead{}          & \colhead{} & 
\colhead{}         & \colhead{}          & \colhead{} &
\colhead{($\times10^{10}$)} & \colhead{($\times10^{8}$)} & 
\colhead{($\times10^{3}$)}  & \colhead{($\times10^{9}$)} & 
\colhead{($\times10^{6}$)}  & \colhead{($\times10^{8}$)} & 
\colhead{($\times10^{8}$)}  & \colhead{($\times10^{6}$)} &
\colhead{} & \colhead{}
}
\startdata
1 &0.9490 &0.0449 &0.0038 &0.0027 & &6.32 &1.62 &2.66 &2.03 &0.86 &1.15 &0.71 &0.73 &1.76 &96.5 \\
2 &0.9461 &0.0472 &0.0040 &0.0031 & &6.30 &1.61 &2.63 &2.13 &0.94 &1.34 &0.85 &0.88 &1.89 &97.6 \\
3 &0.9441 &0.0487 &0.0041 &0.0034 & &6.29 &1.61 &2.61 &2.20 &1.00 &1.47 &0.95 &0.99 &1.98 &98.3 \\
4 &0.9408 &0.0512 &0.0043 &0.0040 & &6.28 &1.60 &2.57 &2.32 &1.10 &1.70 &1.13 &1.18 &2.13 &99.6  \\
5 &0.9343 &0.0585 &0.0049 &0.0027 & &6.25 &1.57 &2.50 &2.65 &1.44 &1.09 &0.77 &0.82 &2.54 &102  \\
6 &0.9288 &0.0630 &0.0053 &0.0033 & &6.23 &1.56 &2.45 &2.85 &1.67 &1.31 &0.96 &1.03 &2.85 &104  \\
7 &0.9250 &0.0661 &0.0056 &0.0037 & &6.21 &1.55 &2.42 &2.99 &1.83 &1.48 &1.11 &1.20 &3.08 &105  \\
8 &0.9204 &0.0697 &0.0059 &0.0044 & &6.19 &1.53 &2.38 &3.16 &2.04 &1.75 &1.33 &1.45 &3.36 &107  \\
9 &0.9233 &0.0690 &0.0058 &0.0024 & &6.21 &1.53 &2.39 &3.12 &2.00 &0.93 &0.71 &0.77 &3.25 &106  \\
10 &0.9180 &0.0733 &0.0062 &0.0029 & &6.19 &1.52 &2.35 &3.32 &2.27 &1.12 &0.87 &0.96 &3.60 &108 \\
11 &0.9143 &0.0764 &0.0064 &0.0032 & &6.17 &1.51 &2.32 &3.46 &2.48 &1.26 &1.00 &1.10 &3.87 &109 \\
12 &0.9080 &0.0815 &0.0069 &0.0040 & &6.14 &1.49 &2.28 &3.70 &2.83 &1.52 &1.23 &1.37 &4.33 &112 \\
13 &0.9142 &0.0779 &0.0066 &0.0016 & &6.17 &1.50 &2.31 &3.53 &2.58 &0.63 &0.50 &0.55 &3.96 &109 \\
14 &0.9076 &0.0837 &0.0070 &0.0020 & &6.14 &1.48 &2.27 &3.79 &2.99 &0.78 &0.63 &0.71 &4.48 &112 \\
15 &0.9027 &0.0879 &0.0074 &0.0023 & &6.12 &1.47 &2.23 &3.99 &3.32 &0.88 &0.73 &0.82 &4.89 &114 \\
16 &0.8946 &0.0949 &0.0080 &0.0029 & &6.09 &1.45 &2.18 &4.31 &3.90 &1.09 &0.92 &1.05 &5.61 &118 \\
17 &0.8911 &0.0911 &0.0077 &0.0106 & &6.06 &1.45 &2.19 &4.13 &3.60 &3.99 &3.35 &3.78 &5.46 &120 \\
18 &0.8793 &0.0992 &0.0084 &0.0136 & &6.00 &1.42 &2.13 &4.50 &4.33 &5.07 &4.35 &4.97 &6.42 &126 \\
19 &0.8708 &0.1048 &0.0088 &0.0160 & &5.96 &1.40 &2.08 &4.76 &4.88 &5.92 &5.16 &5.93 &7.15 &130 \\
20 &0.8595 &0.1116 &0.0094 &0.0199 & &5.90 &1.38 &2.03 &5.07 &5.63 &7.30 &6.48 &7.50 &8.16 &136 \\
\enddata
\tablenotetext{a}{1-16 Non-Standard Solar Models, 17-20 Standard 
Solar Models}
\tablenotetext{b}{Values given in units of cm$^{-2}$s$^{-1}$, except
$\Phi$($^{37}$Cl) and $\Phi$($^{71}$Ga), which are given in SNU.}
\end{deluxetable}
\end{landscape}

\begin{deluxetable}{cccccc}
\tabletypesize{\scriptsize}
\tablecaption{Solar Model Constraint Values\tablenotemark{a}}
\tablewidth{0pt}
\tablehead{
\colhead{Model} & \colhead{R$_{\mathrm {env}}$} & 
\colhead{Y$_{\mathrm {surf}}$} & \colhead{$^{8}$B} & 
\colhead{(Z/X)$_{\mathrm {surf}}$} & \colhead{rms($\delta$c/c)}
}
\startdata
1 &0.7139 &0.1488 &0.86 &0.0183 &0.0123 \\
2 &0.7071 &0.1548 &0.94 &0.0205 &0.0114 \\
3 &0.7026 &0.1588 &1.00 &0.0220 &0.0111 \\
4 &0.6982 &0.1649 &1.10 &0.0244 &0.0106 \\
5 &0.7164 &0.1784 &1.44 &0.0190 &0.0084 \\
6 &0.7102 &0.1876 &1.67 &0.0212 &0.0070 \\
7 &0.7057 &0.1936 &1.83 &0.0228 &0.0063 \\
8 &0.7020 &0.1997 &2.04 &0.0254 &0.0060 \\
9 &0.7185 &0.1967 &2.00 &0.0194 &0.0064 \\
10 &0.7130 &0.2040 &2.27 &0.0217 &0.0052 \\
11 &0.7094 &0.2090 &2.48 &0.0233 &0.0046 \\
12 &0.7061 &0.2166 &2.83 &0.0259 &0.0041 \\
13 &0.7211 &0.2096 &2.58 &0.0197 &0.0056 \\
14 &0.7157 &0.2182 &2.99 &0.0221 &0.0040 \\
15 &0.7124 &0.2242 &3.32 &0.0238 &0.0031 \\
16 &0.7088 &0.2332 &3.90 &0.0265 &0.0022 \\
17 &0.7244 &0.2271 &3.60 &0.0201 &0.0050 \\
18 &0.7199 &0.2375 &4.33 &0.0227 &0.0029 \\
19 &0.7163 &0.2444 &4.88 &0.0244 &0.0015 \\
20 &0.7128 &0.2519 &5.63 &0.0272 &0.0006 \\
\enddata
\tablenotetext{a}{1-16 Non-Standard Solar Models, 17-20 
Standard Solar Models}
\end{deluxetable}

\begin{figure}
\figurenum{1}
\epsscale{0.95}
\plotone{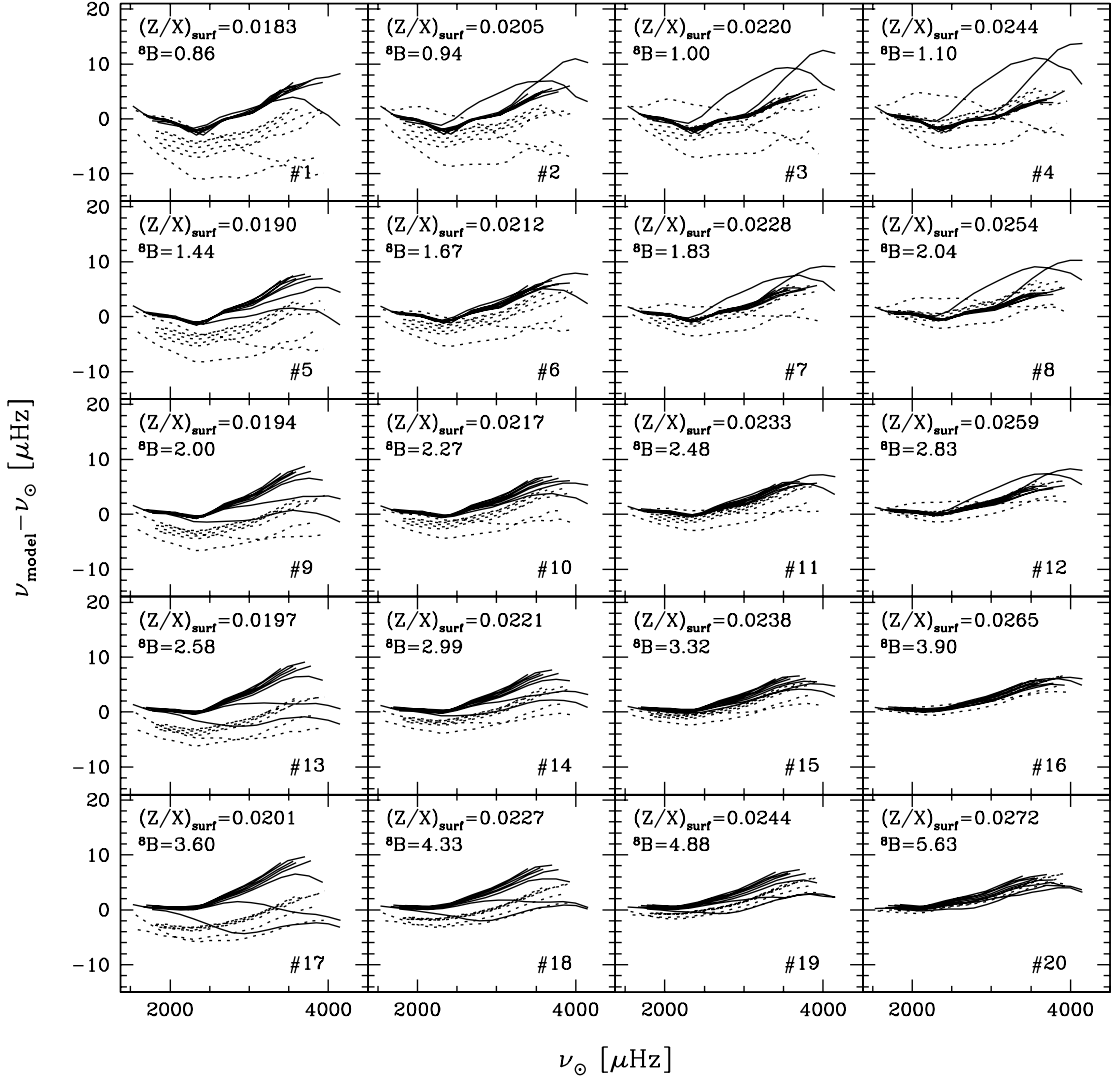}
\caption{Grid of {\it p}-mode frequency difference plots, (model - Sun), 
for all models computed in this work.  Lines connect data with a common
{\it l}-value.  Dashed lines correspond to $l$=0-4, 10 and 20.  Solid
lines correspond to $l$=30, 40, 50, 60, 70, 80, 90, and 100.  Each
plot is annotated with the corresponding model number 
(see Tables 1-3), the surface Z/X ratio, and the $^{8}$B
neutrino flux, in cm$^{-2}$s$^{-1}$.  Observational data used are 
from the MDI-SOHO data set (Rhodes et al. 1997).}
\end{figure}

\begin{figure}
\figurenum{2}
\epsscale{0.5}
\plotone{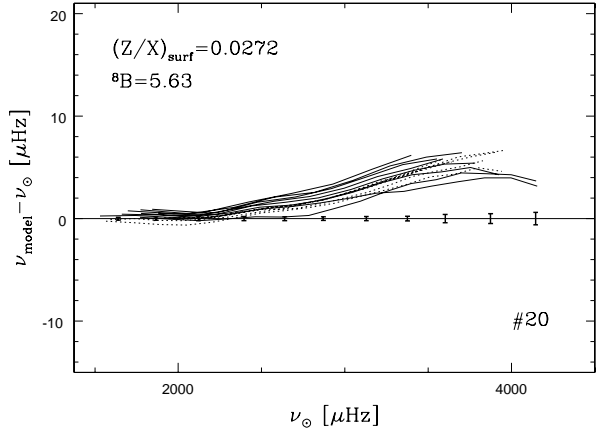}
\caption{{\it p}-mode frequency difference plot, (model - Sun), 
for model \#20 computed in this work.  Lines connect data with a common
{\it l}-value.  Dashed lines correspond to $l$=0-4, 10 and 20.  Solid
lines correspond to $l$=30, 40, 50, 60, 70, 80, 90, and 100.  Listed 
are the surface Z/X ratio and the $^{8}$B neutrino flux, in 
cm$^{-2}$s$^{-1}$.  Observational data used are from the MDI-SOHO data
set (Rhodes et al. 1997).  The error bars indicate 10$\sigma$ errors 
in the data, averaged over 250$\mu$Hz frequency bins, for all 
{\it l}-values included in this work.}
\end{figure}

\begin{figure}
\figurenum{3}
\epsscale{0.9}
\plotone{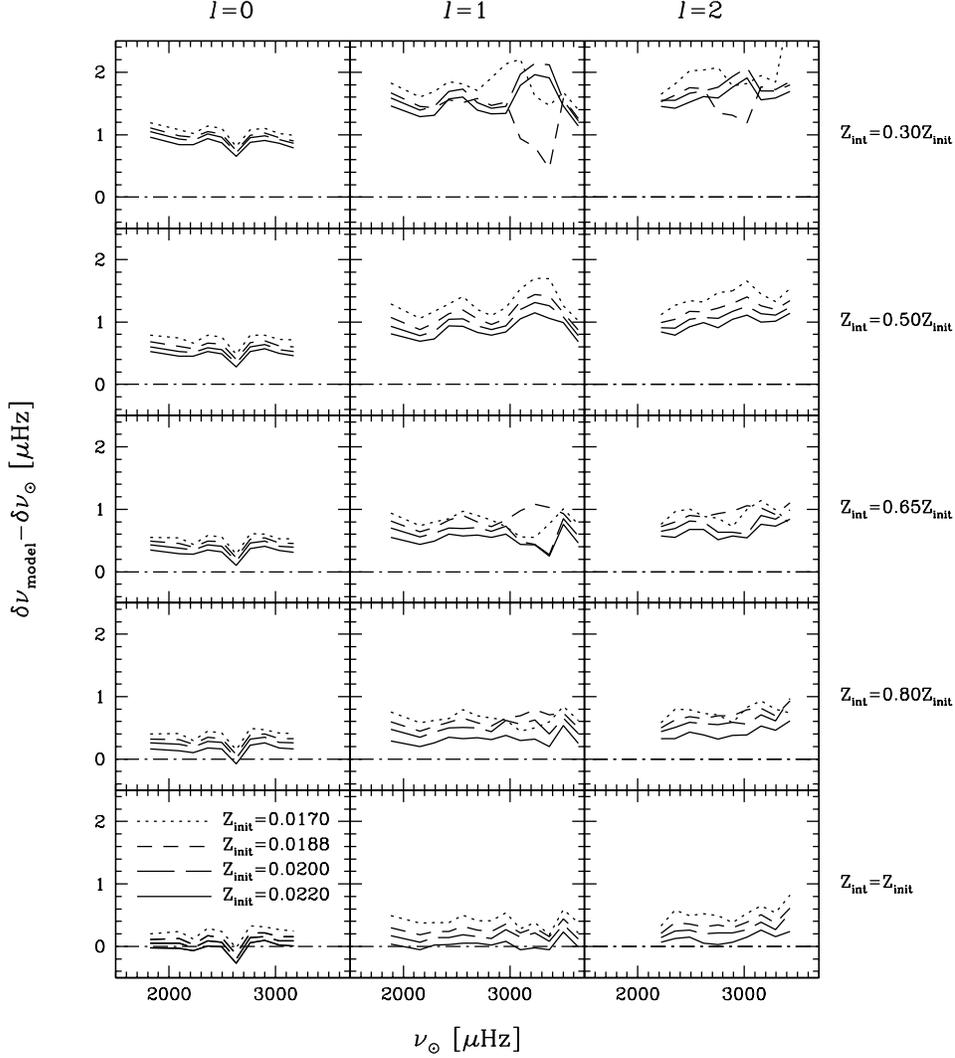}
\caption{Grid of small spacing difference plots, (model - Sun), 
for all models computed in this work.  Lines connect data 
from a particular model, indicated by Z$_{\mathrm {int}}$ and 
Z$_{\mathrm {init}}$.  Each row contains models of the same 
Z$_{\mathrm {int}}$, where Z$_{\mathrm {int}}$=0.30Z$_{\mathrm {init}}$ 
includes models \#1-4, Z$_{\mathrm {int}}$=0.50Z$_{\mathrm {init}}$
includes models \#5-8, Z$_{\mathrm {int}}$=0.65Z$_{\mathrm {init}}$ 
includes models \#9-12, Z$_{\mathrm {int}}$=0.80Z$_{\mathrm {init}}$ 
includes models \#13-16, and Z$_{\mathrm {int}}$=Z$_{\mathrm {init}}$
includes the standard solar models \#17-20.  Each column contains 
results for a common {\it l}-value, with {\it l}=0 in column 1, 
{\it l}=1 in column 2, and {\it l}=2 in column 3.  Observational data 
used are from the MDI-SOHO data set (Rhodes et al. 1997).}
\end{figure}

\begin{figure}
\figurenum{4}
\epsscale{1.0}
\plotone{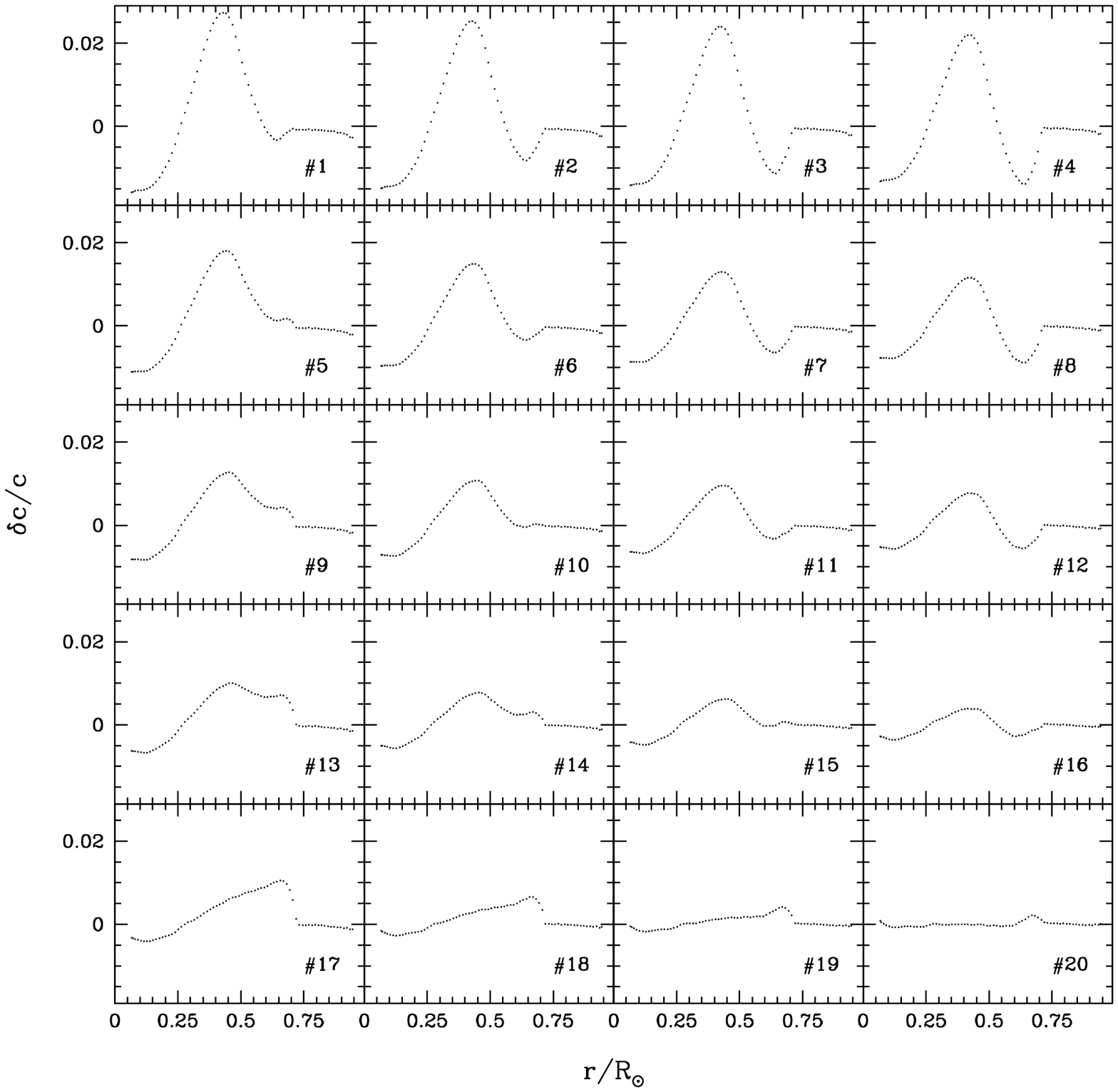}
\caption{Grid of fractional difference plots for the sound-speed  
((Sun - model)/model), for all models computed in this work.  Solar
sound-speed data were determined by Basu, Pinsonneault \& Bahcall 
(2000) using the MDI-SOHO solar frequency data set 
(Rhodes et al. 1997).}
\end{figure}

\begin{figure}
\figurenum{5}
\epsscale{1.0}
\plotone{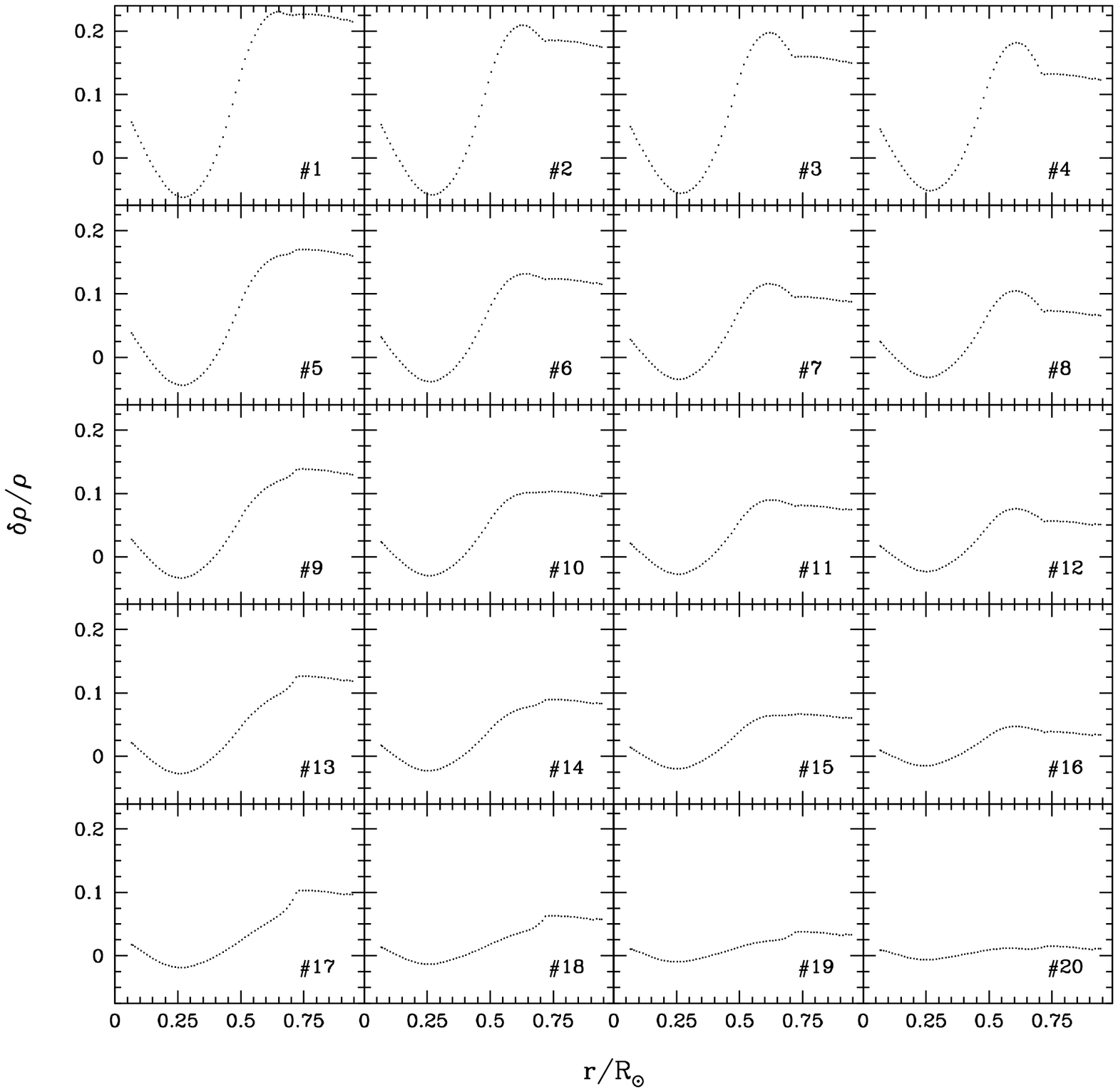}
\caption{Grid of fractional difference plots for density 
((Sun - model)/model), for all models computed in this work.  Solar
density data were determined by Basu, Pinsonneault \& Bahcall (2000) 
using the MDI-SOHO solar frequency data set (Rhodes et al. 1997).}
\end{figure}

\begin{figure}
\figurenum{6}
\epsscale{0.5}
\plotone{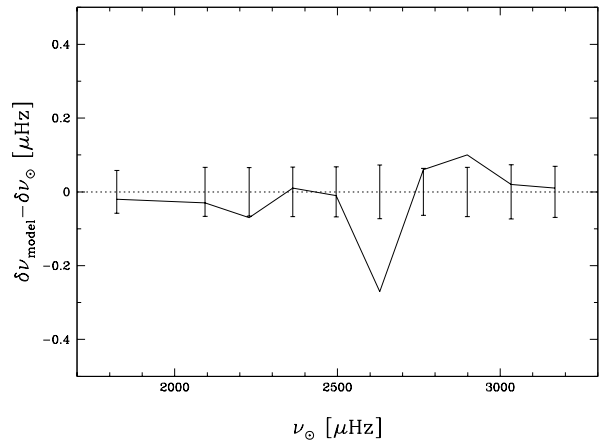}
\caption{Small spacing difference plot (model - Sun), 
for standard solar model \#20 with $l$=0.  Observational 
data used are from the MDI-SOHO data set (Rhodes et al. 1997).  
The error bars plotted represent errors in the data.  The 
observational uncertainties are approximately $\pm$0.07 $\mu$Hz
for the small spacing determination.}
\end{figure}

\begin{figure}
\figurenum{7}
\epsscale{0.5}
\plotone{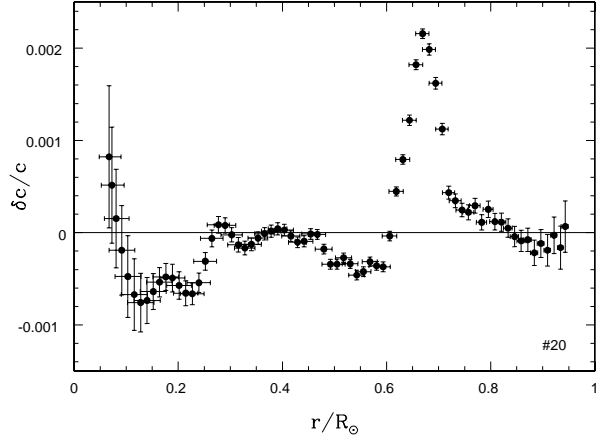}
\caption{Fractional difference plot for the sound-speed 
((Sun - model)/model), for standard solar model \#20 (zoom in of 
Figure 4).  Vertical error bars indicate 1$\sigma$ errors in the 
inversion results due to errors in the data.  Horizontal error
bars are a measure of the resolution of the inversion.}
\end{figure}

\begin{figure}
\figurenum{8}
\epsscale{0.5}
\plotone{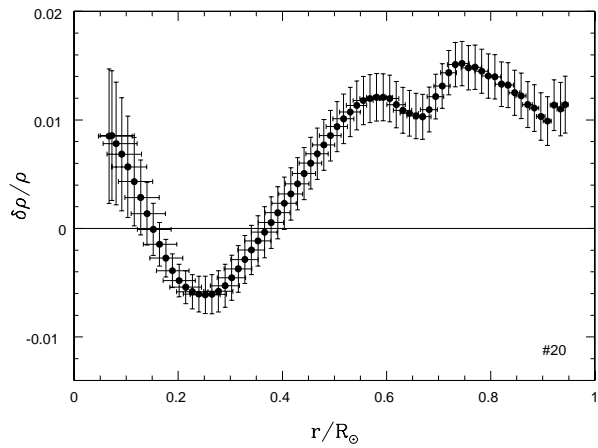}
\caption{Fractional difference plot for density 
((Sun - model)/model), for standard solar model \#20 (zoom in of 
Figure 5).}
\end{figure}

\end{document}